\documentclass[
reprint, aps, prb
]{revtex4-2}

\usepackage{graphicx}
\usepackage{dcolumn}
\usepackage{bm}
\usepackage{textcase}

\begin{document}
\title{Control of Genes by Self-organizing Multicellular Interaction Networks}

\author{Kyle R. Allison}
\email{kyle.r.allison.phd@gmail.com}

\affiliation{ Unaffiliated, Atlanta, USA; Previously: Department of Medicine, Division of Infectious Diseases, Emory University School of Medicine, Atlanta, GA }

\begin{abstract}
\noindent Multicellular self-organization drives development in biological organisms, yet a comprehensive theory is lacking as basic properties of cells can complicate common approaches. Framing such properties by dynamic graphs led to new theoretical propositions for multicellular self-organization in \textit{Escherichia coli}. Here, corresponding ideas are developed from biologically-general first principles. The resulting perspective could aid both experimental and computational approaches to multicellular biology as well as efforts to control and engineer it.
\end{abstract}

\maketitle

\noindent The following is a unique perspective on multicellular self-organization which may be useful to developing a general theory of the subject. The central ideas originated from exploration \cite{puri_multicellular_2025, allison_multicellular_2025} of multicellular self-organization in \textit{Escherichia coli} \cite{puri_evidence_2023, puri_fluorescence-based_2023, puri_escherichia_2024}, representing a special case which occurs in a unicellular organism. Despite this particular origin, these ideas have generality and may be broadly applicable to any organism that employs multicellular self-organization. They can also be arrived at generally—by sparsely drawing key concepts from several subjects, including self-organization, cell biology, and graph theory—rather than solely from \textit{E. coli} cell- and molecular biology. A comprehensive review of multicellular self-organization theory, illustrated via a diverse collection of biological examples, will likely become more practical in the future.

\section*{\NoCaseChange{Self-organization}}
\noindent Self-organization characterizes many natural processes spanning physical and temporal scales, from the microtubules that structure eukaryotic cells to the hierarchical social networks generated among animals in populations. Self-organization describes when discernible patterns arise due to the dynamic interactions between a system’s individual components, rather than being externally imposed or centrally specified such as by circuit wiring diagrams or architectural blueprints. The “self” indicates organization comes from inside the system, rather than outside, though all systems have external constraints and non-equilibrium thermodynamics indicates self-organization requires the internal dissipation of energy from external sources \cite{nicolis_self-organization_1977}. Considering dynamics, the interactions and patterns comprising self-organization are semi-stable, differing from self-assembly where they are considered fully stable \cite{halley_consistent_2008}. Whereas self-assembly spontaneously tends toward structured equilibrium states (\textit{e.g.}, nanoparticle assembly), self-organization instead transitions from one semi-stable attractor state to another in a repeatable sequence, performing a choreographed dance through dynamic phase space (one might choose to  visualize this mathematically as a series of semi-stable basins of attraction that empty one into the next). Empirically, self-organization and self-assembly sometimes appear indistinguishable, but they have incompatible dynamic implications: self-assembled patterns are inevitable and persist in the absence of substantial perturbation; self-organized patterns on the other hand are contingent on circumstances and inevitably disappear. Self-organization is then, more narrowly, when dynamic interactions between components create repeatable sequences of semi-stable patterns. The duration of semi-stable states can vary, for example, morphological transitions in embryogenesis can occur in minutes while other transitions in adult tissues can take decades. Such differences in characteristic timescales mean some semi-stable states within a self-organizing sequence have “temporal stability,” in a relative sense if not in a strictly dynamic sense. Semi-stability requires feedback loops, where the change in a variable (\textit{e.g.}, a component’s properties) depends on its current state (\textit{e.g.}, its current properties). Consistently, positive- and negative feedback loops are common mechanisms of self-organization, but they can be obscured when mediated via interactions among multiple components. Emergence, where global properties “emerge” from interactions between components rather than strictly from the components’ individual properties, results from self-organization. This idea is at times viewed as more epistemological than scientific, and whether a given pattern is thought to “emerge” from or be dictated by a system’s components usually reflects the degree of mechanistic understanding.

Self-organization’s origin as a concept was philosophical, “\textit{Selbstorganisation}” being coined in the 18th-century by Immanuel Kant in his \textit{Critique of Judgment} to distinguish living things from machines. It is then ironic that many of the pioneering scientific theories of self-organization arose form complex non-living systems, like chemical networks and machines \cite{nicolis_self-organization_1977}. These contributions applied concepts from thermodynamics and information theory \cite{heylighen_complexity_2009, haken_information_2017}, corresponding to the particular fundamental components of those systems. However, self-organization’s dependence on interactions and their governing frameworks suggests that as a theory becomes more accurate for one class of self-organization, it becomes less applicable to others arising from different interactions with incompatible underlying principles. Rather than detract from theoretical approaches, this illustrates any theory of self-organization is explicitly circumscribed and that each form of self-organization needs its own theory based on the appropriate underlying principles. Or as Kant might have put it: in theories of self-organization, it may be necessary to reject the general in order to make room for the particular.

\section*{\NoCaseChange{Biological Self-organization}}
\noindent Self-organization defines biology at all scales, from the ecological down to the molecular. Protein folding can largely reflect self-assembly of equilibrium states \cite{dobson_protein_2003}, but the dynamic interactions of proteins with one another tend to result from self-organization. At the opposite end of the spectrum, ecological populations can be modeled through stable, equilibrium states, but their underlying adaptive processes are usually recognized as more complicated \cite{levin_self-organization_2005}. The most substantial contributions to biological self-organization perhaps have resulted from studying the internal molecular structures of eukaryotic cell biology \cite{misteli_concept_2001, karsenti_self-organization_2008, mitchison_self-organization_2021}, with microtubules serving as a platform for pioneering work \cite{heald_self-organization_1996, desai_microtubule_1997, nedelec_self-organization_1997}. Molecules and proteins are the fundamental units here, and intracellular examples of molecular self-organization, like cytoskeletal formation and biomolecular phase separation \cite{banani_biomolecular_2017}, are largely understood through chemical and thermodynamic principles. On the other hand, the principles of how cells diversely interact are less apparent, and a general theory of multicellular self-organization remains lacking (though specific cases have been ingeniously addressed \cite{turing_chemical_1952}). This is despite its inherent significance to developmental biology \cite{schweisguth_self-organization_2019} and proximity to Kant’s original idea.

Multicellular self-organization drives development of tissues and organs, and deciphering its molecular events and interactions is a key part of developmental biology \cite{dos_santos_single_2019}. Separately, treating living organisms as “active matter” has enabled the estimation of important mechanical properties of tissues (\textit{e.g.}, elasticity or viscosity)\cite{ramaswamy_mechanics_2010, shankar_topological_2022}. Nevertheless, physics-first models have limited explanatory power for multicellular self-organization \cite{youk_discreteness_2026}, in part because they often omit core cellular properties governing interaction dynamics. Promisingly, synthetic biology and tissue engineering approaches to program development, organoids, self-organization, \textit{etc.} have been uncovering empirical rules for multicellular self-organization \cite{johnson_engineering_2017, toda_programming_2018, ebrahimkhani_synthetic_2019, brassard_engineering_2019, mizuno_robust_2024}. In the future, such engineering approaches would benefit from a constraint-based theoretical framework that enables predictions reliable enough to reduce the current need for trial-and-error validations.

A theory rendering multicellular self-organization predictable has not yet been developed, though one must necessarily exist. The deep reliability with which single zygote cells develop into functional, complex multicellular organisms requires such a framework. Categorically, the rules for such a theory must derive from the properties of its fundamental units: cells. \textit{A priori} and as alluded to above, applying principles from other theories of self-organization (\textit{e.g.}, thermodynamics or ecological laws) to multicellular self-organization will typically be inadequate: their basic units, properties, and interactions are not equivalent. Of course, certain dynamical features are definitional to self-organization (semi-stability, positive- and negative feedback loops, \textit{etc.}), and some parallels between different classes of self-organization are expected. A model for one system may be made to approximate another by tuning its parameters, but its differing implications may lead to misinterpretations.

\section*{\NoCaseChange{Multicellular Self-organization}}
\noindent Across the tree of life, cells perform diverse and myriad behaviors \cite{sebe-pedros_biodiversity_2025}, the details of which are essential to the multicellular self-organization of each species and the differences between species. A general theory of multicellular self-organization, or multicellular organization more broadly, must be grounded in properties common to all cells and how such properties influence one another’s dynamics. Two such properties are division and adaptation (\textit{i.e.}, gene regulation, not the statistical, population phenomenon of evolution): cells divide and they adapt (\textbf{Fig. 1}). These traits are non-trivial and tend to distinguish “multicellular” self-organization from other forms. Division means new interacting units (cells) are regularly produced from molecular components at a lower scale. Effectively, this introduces new interacting units (cells) into the system \textit{de novo} at the scale of interactions, posing challenges for equations-based approaches that can already be tricky with a constant number of self-organizing components. Adaptation further complicates things: cell properties and interaction rules change with time and context, meaning they are temporally local (\textit{i.e.}, time-specific) and are expected to differ from one stage to the next. The complications arising from these basic properties of cell biology may help illustrate why a theory of multicellular self-organization is still missing. Fortunately, biology as it exists has supplied a useful constraint.

\begin{figure}
\includegraphics{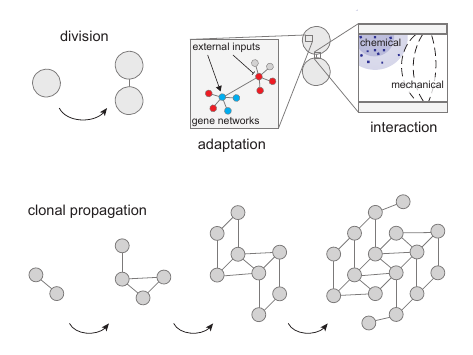}
\caption{Dynamic graphs for multicellular self-organization.
Cells divide (top left). Cells can adapt their identity and behavior by regulating genes using gene networks that interpret molecular signals, including inputs from their external environment (top right). Proximal cells can form chemical and mechanical interactions and can modify these interactions by adaptation (top right). Self-organization is determined by the specific interactions among a system’s fundamental units. Cells are the fundamental units of multicellular self-organization. Dynamic graph models, which represent the units and interactions of dynamically changing systems, are a logical framework for multicellular self-organization (and more generally, for any form of self-organization). In this framework, cells are represented as nodes and interactions as edges (bottom). As cells divide, adapt, and interact, new nodes and edges are added to a propagating dynamic graph. Clonality (where all cells descend from a single progenitor cell) simplifies propagation: all changes to nodes and edges result from division, adaptation, and interaction events on preceding graph stages, rather than from unpredictable external effects.}
\end{figure}

All complex multicellularity known to date results from clonal multicellular self-organization, where the entire organism develops from an individual cell \cite{knoll_multiple_2011}. Arguments based in evolutionary theory have been made for this (\textit{e.g.}, conflict avoidance), though an independent argument can be made by illustrating the limits on non-clonal organization from a graph-based theory of multicellularity \cite{puri_multicellular_2025}. While non-clonal complexity remains possible, clonality covers known complexity and advantageously simplifies self-organization. Despite the difficulties of division and adaptation, clonality entails all new units and interactions result from those already present and not unpredictably from outside.

Multicellular self-organization then begins equivalently in all cases. A single cell divides into two connected cells which can interact, even if only briefly. These two must choose whether to remain adhered or to separate, and most organisms have molecular and behavioral mechanisms to achieve either option. If they remain together, they must choose how to interact: cooperate, coordinate, communicate, compete, or kill each other; reposition or change shape to accommodate one another; differentiate and adopt complementary roles; or ignore each other. There are many possible decisions, but time is limited because the pair will soon divide and, as four cells, face a similar set of questions in a new context shaped by earlier responses. This iterative cycle—divide, interact, adapt—repeats until cells stop dividing, by choice or circumstance, and either reach a state of terminal differentiation or separate from each other. At every stage, cells make decisions framed by their past decisions and neighboring cells, and individual adaptions can dramatically alter potential future interactions.

The contingency of such cycles indicates the principles underlying each interaction-adaptation connection can change and are temporally local. For example, an equation-based model capturing cell-cell adherence at one stage will be superseded or replaced by collective generation of extracellular matrices at a successive stage. Cells both alter and respond to their external environment, meaning cells growing closely in communities inevitably interact, even if such interactions are not always strictly considered cell-cell signals. How cells adapt to interactions and external inputs is a result of their evolutionary history and largely unpredictable purely from chemical or mechanical principles. However, clonality means each sequence of divide-interact-adapt stages are causally connected and their dynamics can be separated into modules and investigated with a degree of independence. For example, organ development can be studied from a key set of interacting cells without needing to track the entirety of development from the first cell. This modularity of self-organization also explains the success of organoid models that recapitulate organ physiology without originating by natural development from a single zygote.

Alan Turing’s landmark reaction-diffusion model for pattern formation in tissues remains highly influential for studies of multicellular self-organization \cite{turing_chemical_1952}. It demonstrated simple chemical rules could lead to complex patterns in chemical gradients and was later successfully applied to a variety of periodic patterns in tissues such as zebras’ stripes or the formation of teeth and fingers. To render the mathematics more tractable, core biological properties were omitted, including division, gene regulation, and mechanical interactions.  Hence, the model naturally has specific strengths and limitations for multicellular self-organization \cite{landge_pattern_2020}, and while it brilliantly illustrates how periodic patterns can arise from chemical interactions, it is not a comprehensive theory of morphogenesis. Similarly, while equations-based models can accurately explain individual types of interactions, they can only capture parts rather than the whole of multicellular self-organization.

\section*{\NoCaseChange{Dynamic Graphs for Multicellular Self-organization}}
\noindent Dynamic graph models can capture individual components along with their changing interactions, providing a useful framework for multicellular self-organization and its spatially disparate, adapting cell interactions which are governed by an assortment of underlying principles. Dynamical graph models were first proposed as general theoretical framework for applied mathematics and computation \cite{harary_dynamic_1997}. Unlike static graphs, whose components and interactions are fixed, dynamic graphs have nodes and edges that change (appearing, disappearing, adapting, \textit{etc.}). Whereas static graphs can be understood by properties like connectivity, clustering, path lengths, \textit{etc.}, dynamic graphs are dominated by their rules of propagation and can often be understood as a causal sequence of individual static graphs. Often interchangeable with "temporal networks," they have critical advantages over more classical static networks, as previously discussed for several applications \cite{holme_temporal_2012} and multicellular self-organization in particular \cite{puri_multicellular_2025}. Dynamic graphs have been widely applied for causal reasoning and machine inference and can be effectively learned by neural-network algorithms \cite{pareja_evolvegcn_2019}, but basic biologic properties have largely been absent, cell division for example only recently being incorporated into such approaches \cite{yamamoto_probing_2022}. Dynamic graphs were first employed to theorize multicellular self-organization, separate from computation, against the empirical backdrop of \textit{E. coli} multicellular dynamics \cite{puri_multicellular_2025}. Potential principles of multicellular self-organization were then derived from this basis by exploring simple model examples.

Within this framework, nodes are cells and edges are the chemical and mechanical interactions between them. Division creates new nodes and new interactions (\textbf{Fig. 1}). Interactions between cells are also influenced by adaptation through gene expression. Adaptation in turn effects the particular interactions each cell experiences at any given stage, thereby creating potentially vast webs of feedback loops between adaptation and interaction \cite{allison_multicellular_2025}. As division is discrete and its characteristic time scale is comparable to gene expression in many organisms, a degree of self-organization can be modeled as discrete-time dynamic graphs with Boolean rules for adaptation and graph propagation (see also argument for \textit{E. coli} \cite{allison_multicellular_2025}). Although chemical and mechanical signaling can be fast and noisy, their magnitudes and rates often allow time averaging over windows like the cell division cycle to produce robust non-noisy genetic decisions. Discussed more below, multiple types of interactions can exist between cells, many of them local (\textit{e.g.}, direct contact) but others are more global (\textit{e.g.}, morphogen signaling). As cells divide, the multicellular graph grows and along with it a network of multicellular interactions. Furthermore, as communities branch and diversify during morphogenesis, so to do their multicellular interaction networks which become increasingly heterogeneous and distributed.

Though this framing adds a dimension of complexity, dynamic graphs of clonal self-organization can usually be sketched with accuracy by hand if the modeled processes have been empirically observed by multicellular dynamics (see examples in \cite{puri_multicellular_2025}). Like multicellular self-organization, dynamic graphs are modular meaning individual subgraphs of smaller sets of cells can be analyzed in isolation from most long-range interactions and the overall graph. Hence, even in complex organisms, localized simple cases of multicellular self-organization (\textit{e.g.}, within tissue development or organogenesis) can be separately considered and sketched. Dynamic graph models also lend themselves to computation, and there may be considerable potential for human health in applying appropriate cellular constraints to construct dynamic graph models for multicellular self-organization. As an illustrative example, an attention-based algorithm to uncover dynamic graphs was recently used to learn the connection between specific tumor niche architectures and clinical outcomes \cite{zuo_stclinic_2025}. Such studies indicate more-predictive dynamic-geometry-based prognoses are on the horizon and contribute to a shift in understanding some cancers as aberrant multicellular self-organization, resulting from mis-regulation of normal genetic programs, rather than selfish evolution of individual cells.

\section*{\NoCaseChange{Multicellular Interaction Networks}}
\noindent Dynamic multicellular interaction networks (\textbf{Fig. 2}) are a logical consequence of framing multicellular self-organization by dynamic graphs \cite{allison_multicellular_2025}. Such networks arise because cells in communities impose chemical and mechanical effects on one another (electrical, magnetic, and thermal effects may also be usefully considered in cases). Interactions are inevitable: cells exert forces and alter the chemical properties of their shared environment, producing combinatorial sets of simultaneously experienced interactions. Division creates spatial asymmetries in communities, and many organisms perform asymmetric division intentionally to diversify adaptation among sister cells. Further division can amplify the diversity of cell adaptations and interactions iteratively at each stage of self-organization. This entails interaction networks are spatially distributed and non-uniform. As mentioned above, interactions between cells can be interpreted by gene networks to drive adaptation. Hence, multicellular interaction networks are transient webs of locally specified gene-network inputs covering self-organizing populations of cells. 

\begin{figure}
\includegraphics{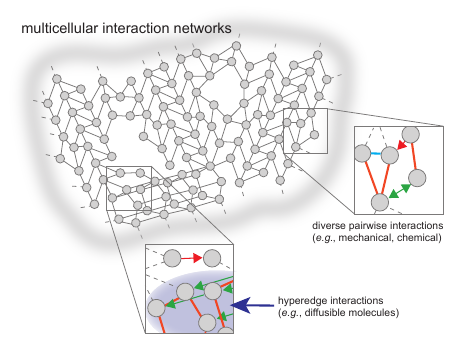}
\caption{Multicellular interaction networks.
Interactions (edges) between cells (nodes) dynamically change and are distributed over populations of self-organizing cells. The set of such interactions in the dynamic-graph framework represents a multicellular interaction network. The individual edges of such networks can be directed, weighted, combinatorial, and/or diverse (or not). Many mechanical and some chemical interactions between cells can be encoded by such pairwise edges. However, some multicellular interactions, like those mediated by diffusible molecules (\textit{e.g.}, morphogens, metabolites, and so on), simultaneously connect many cells. These interactions would be better encoded as hyperedges which simultaneously connect many nodes. Hyperedges were developed to address the pair-wise limitation of classical graph theory for complex systems.}
\end{figure}

Pair-wise edges between nodes are largely effective for capturing direct mechanical interactions between cells. They can be undirected to capture reciprocal interactions or directional when forces between cells are sensed by one but not the other. Chemical interactions mediated by diffusible molecules are more diffuse and might be poorly captured by pair-wise edges (neuronal synapses being an important exception). Diffusible chemical signals, like morphogens and quorum-sensing molecules, would be better represented as hyperedges which connect sets of nodes and address a common limitation within classical graph theory. The incorporation of hyperedges into graphical approaches has been elaborated for intracellular networks \cite{klamt_hypergraphs_2009}, and similar ideas would apply to multicellular interaction networks. Hyperedges also enable encoding of multicellular interactions based on the depletion or absence of diffusible molecules or collective mechanical properties of tissues studied, both of which are regularly gene-network inputs.

To be self-organizing, the semi-stable patterns of multicellular interaction networks must occur in a reliable sequence. Such interactions can be critical inputs for gene networks thereby determining adaptation and future multicellular interactions. Thus, for cell adaptations to occur reliably at a developmental stage, gene network inputs and corresponding multicellular interactions must also be reliably generated. Correspondingly, dynamic geometries and properties that enable the reliability of multicellular interaction networks are central to multicellularity and development. From this constraint, several ideas follow naturally, like the importance of forming developmental rosettes or extending internal volumes with controlled dimensions \cite{puri_multicellular_2025}. Many potential multicellular interactions can coincide, providing cells an opportunity to encode spatial position and developmental time by combinatorially regulating genes through multiple input signals. Such combinatorial regulation would be an effective way to enable robust gene expression. Alone, a chemical or mechanical signal might just be a transient property of the environment, but in conjunction such signals could uniquely predict a cell’s specific multicellular position and stage. Such combinatorial regulation by multicellular interactions networks provided a useful explanation for the robust and sequential activation of otherwise-noisy genes during \textit{E. coli} self-organization \cite{allison_multicellular_2025}.

\section*{\NoCaseChange{Control of Genes by Multicellular Interaction Networks}}
\noindent While multicellular interaction networks can serve as inputs for gene networks, they are also produced as the outputs of gene networks, raising the potential for dynamically-changing webs of feedback loops between cell adaptation and interaction. In clonal communities, the types of interactions and their degree, location, timing, and combination with other interactions all directly result from divisions and adaptations of preceding cells. Given the chain of causality for clonally-organizing cells, progeny share substantial identity with progenitors from which they have divided and diversified. In the clonal case then, cells can specify the inputs for their own gene networks and use multicellular interaction networks to control their own genes, previously introduced as multicellular control of gene networks \cite{allison_multicellular_2025}. “Control,” in this engineering sense, means robustly determining a variable’s dynamic changes, and distinguishes this phenomenon from more straightforward gene induction or gene regulation. This “multicellular control” of genes by self-generated interaction networks follows logically from the division, adaptation, and interaction of cells in clonal communities when considered within a graphical framework.

Multicellular interaction networks may therefore function as a dynamic and distributed control layer for gene networks in clonally self-organizing communities, thereby precisely determining their sequential adaptations. This control mechanism, by contrast, is not readily available in non-clonal organization, where multicellular interactions are probabilistic and lack predictability. Multicellular control also illustrates a mechanism for iteratively evolving multicellular complexity: re-configuring gene networks to control beneficial traits via the stage-dependent multicellular interactions generated by a community \cite{allison_multicellular_2025}. As a corollary, self-organizing multicellular interaction networks will in cases be essential for explaining the complicated genetic circuitry within cells, as previously suggested \cite{allison_multicellular_2025}. Multicellular control of genes by interaction networks allows cells in clonally self-organizing communities a unique way to control their own genes, behaviors, and fates. Further consideration in the context of community multicellular dynamics reveals the importance of multicellular control for self-organization.

\section*{\NoCaseChange{Multicellular- and Developmental Daisy Chains}}
\noindent Reliable sequences of semi-stable patterns are a primary property of self-organization. Multicellular control by interaction networks would directly enable this property. Gene-network outputs at one stage generate multicellular interactions that then serve as inputs for gene networks at later stages, \textit{i.e.}, gene networks are temporally \textit{daisy-chained} through multicellular interaction networks \cite{allison_multicellular_2025}. Looking forward in time, gene networks specify the inputs that will be experienced by cells at future states; looking backward, the inputs for key gene networks have been reliably produced by the genetically regulated behaviors of preceding cells. The iterative coupling of interaction-adaptation feedback loops through multicellular control thereby constitutes a “multicellular daisy chain” for a clonal community’s dynamics (\textbf{Fig. 3}). First introduced for understanding \textit{E. coli} gene regulation during self-organization \cite{allison_multicellular_2025}, this model is also derivable from the basic properties of cells. “Developmental daisy chains” is then a more appropriate description for obligate multicellular organisms whose developmental behaviors are widely accepted. Within daisy chains, multicellular interaction networks serve as a dynamic and distributed control layer, and the corresponding gene networks serve as the hard-wired logic for sensing and modifying this control layer. To represent daisy chains by dynamic graphs, interactions are the edges (and hyperedges) that change as graphs propagate and gene networks are a separate form of static graphs that represent how nodes interpret these edges and adapt them to define the rules of graph propagation. As interaction networks branch and diversify during morphogenesis so would the daisy chains they generate. In this way, the key inputs determining the behavior of each cell at each time can be reliably specified within the interaction-adaptation logic of a daisy chain.

\begin{figure}
\includegraphics{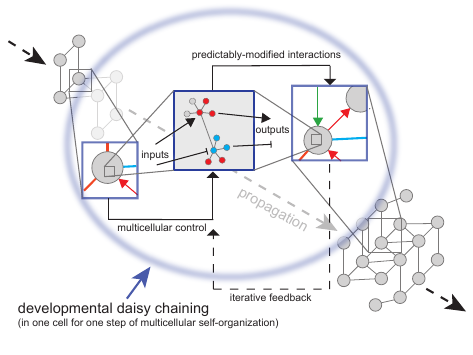}
\caption{Developmental daisy chains.
As groups of cells propagate by clonal multicellular self-organization (represented as the transition from the graph in the upper left to the one in bottom right), each cell experiences specific combinations of interactions (inset, left) deriving from the communities’ overall multicellular interaction network (\textbf{Fig. 2}). These interactions can serve as inputs for each cell’s gene networks (inset, middle) via multicellular control, as articulated. The outputs of these gene networks drive adaptation, including cell-fate and -differentiation decisions. Moreover, these outputs can modify the cell’s local multicellular interaction network (inset, right). In clonal populations, such modifications are predictable because they were generated by progenitors and their neighbors. Modified interactions feedback iteratively as dynamic and distributed inputs for subsequent rounds of multicellular control. Such predictive and iterative feedback of gene networks and multicellular interactions networks in clonal self-organization has been described as multicellular daisy chaining. The resulting multicellular- or developmental daisy chains provide an underlying mechanism for clonal multicellular self-organization by linking the dynamics of cell division, adaptation, and interaction. Moreover, they model how clonal self-organizing populations can robustly and adaptively control the fate, differentiation, and morphogenesis of each of their cells.}
\end{figure}

Developmental daisy chains provide a model for how a series of semi-stable multicellular patterns are reliably propagated by a clonal community based on the adaptations and interactions of cells, their fundamental units. Without this sequential coupling, multicellular interactions and corresponding gene-network inputs would be unreliable. Cells’ inputs would lack predictability and as a result their adaptations and behaviors would as well. Considerations discussed above and previously \cite{allison_multicellular_2025}, point to the possibility that daisy chains are necessary and sufficient for producing multicellular self-organization (a more detailed argument could be made in the future). This would entail that every functional daisy chain creates self-organizing behavior, and any instance of multicellular self-organization relies in part on a daisy chain. It further suggests that any new form of multicellular self-organization (evolved or programmed by synthetic biology) would require creation of a new daisy chain, either \textit{de novo} or by modifying an existing daisy chain (achievable by rewiring its gene networks or controlling it through its external inputs). Though here presented theoretically, the daisy chain model is supported by interpretation \cite{puri_multicellular_2025, allison_multicellular_2025} of the detailed evidence of \textit{E. coli} multicellular self-organization \cite{puri_evidence_2023,puri_fluorescence-based_2023, puri_escherichia_2024}.

Multicellular self-organization is one among many cellular phenomena critical to developmental biology. Multicellular control and daisy chains complement existing gene regulation paradigms like homeostasis and autonomous development (\textit{i.e.}, “developmental gene cascades” and “hierarchical gene regulatory networks”). While homeostatic regulation maintains critical cellular processes through feedback loops, autonomous developmental regulation drives cell-intrinsic programs that operate independently of external cues. Multicellular control adds to these by using self-produced multicellular interactions as dynamic inputs to coordinate stage-specific adaptations. While each example of development may result from its own complicated and varied combination of these paradigms, multicellular control uniquely explains the underlying processes of multicellular self-organization. In many sufficiently large and diversified multicellular communities, separate subgroups of cells (\textit{i.e.}, tissues and organs) will take on differing behaviors and inevitably interact with each other as individual components of a higher scale of self-organization. Similar points regarding predictable dynamics and reliable propagation would apply to such tissue-tissue interactions, and parallel daisy-chain-like ideas could illustrate how networks of tissue interactions serve as predictable inputs for determining dynamic multicellular interaction networks at the more granular scale.

As the complexity of a daisy chain would grow in proportion to the number of divisions, adaptations, and branching points, computational approaches which excel at keeping track of many contingent and combinatorial interactions will be useful. As discussed \cite{puri_multicellular_2025} however, some multicellular geometries (or sub-graphs) are better than others at enabling reliable propagation of interaction networks. Complex daisy chains can therefore be refactored into modular sets of such multicellular interaction motifs \cite{allison_multicellular_2025} which can be studied individually. As multicellular interactions provide the rules for multicellular self-organization, multicellular interaction motifs would be foundational to multicellular biology.

The idea that cell behaviors are determined by networks of interacting molecular components launched systems biology and inspired many computational and machine-inference approaches. It also suggested that deterministic models might be a remote possibility given cells’ internal complexity. Modular cell biology \cite{hartwell_molecular_1999}, where the behavior of individual sub-networks or “modules” (\textit{e.g.}, signaling cascades) could be deciphered by precise theory-guided experiments, suggested a way forward and motivated a generation of experimental systems biologists. Analogously, a \textit{modular multicellular biology} to investigate dynamic multicellular interaction networks and their connections to gene networks could motivate a new generation of systems biologists to elucidate foundational multicellular motifs and put them together to comprehensively understand multicellular self-organization and predictably program it. \textit{Modular multicellular biology}, defined as such, may be essential for understanding multicellular organisms. Enticingly, multicellular interaction networks are both spatially and temporally modular and are sparser than molecular interactions networks: theoretically, modular multicellular biology should prove more tractable than modular cell biology has been, even if the necessary experiments may be challenging.

More immediately, this perspective could help reframe multicellular bacterial infections, a persistently urgent set of problems. Many infections are characterized by multicellular bacterial communities, comprising dozens to hundreds of cells. Even without acquired antibiotic resistance, such communities are highly tolerant to antibiotics and contribute to persistent infections (in turn driving greater antibiotic use and the spread of resistance). \textit{Mycobacterium tuberculosis} has plagued humans as a leading cause of death worldwide for centuries. \textit{M. tuberculosis} forms clonal multicellular chains (known since the microbe’s discovery and termed “cords”) that have biofilm-like properties and were recently documented in samples from patients \cite{lerner_mycobacterium_2020}. These chains are similar to those \textit{E. coli} self-organizes \cite{puri_evidence_2023, puri_escherichia_2024} and also evidently create and extend an internal space between cells. More direct studies of the multicellular dynamics of \textit{M. tuberculosis} cords might reveal they do indeed self-organize by feedback between multicellular interactions and gene expression and perhaps drive antibiotic-tolerant dormancy like in \textit{E. coli}. Such findings would indicate that targeting \textit{M. tuberculosis} multicellular interactions could improve existing antibiotics and halt the behaviors that make this microbe so deadly.
\vspace{4 mm}

\section*{\NoCaseChange{Outlook}}
Since Antonie van Leeuwenhoek’s improvements to microscopy centuries ago, we have looked at cells and seen what they do. But to see patterns, especially ones that dynamically morph and disappear, we must know what to look for, and theoretical models have regularly guided empirical study of cell behavior. The ideas presented here, \textit{e.g.} multicellular interaction networks, multicellular control, developmental daisy chains, \textit{etc.}, can help frame observations and interpretations in diverse organisms, and there may be little value in additional purely theoretical developments. Moreover, their basis in dynamic graphs also suggests these ideas can be adapted into computational approaches to uncover the governing interactions of multicellular organisms, which are composed of many individual cells that divide, adapt, and interact in diverse, distributed, dynamic ways. Future advances will certainly uncover additional self-organizing multicellular patterns common across differing organisms and possibly foundational interaction-network motifs for self-organization. More practically though, they will be integral to reverse engineer diseases and forward engineer new biological solutions.

\vspace{8 mm}
\noindent \textbf{ACKNOWLEDGEMENTS} \\
\noindent The author thanks the two reviewers for their comments which have thoroughly guided revisions.  No funding was received for this research.

\vspace{4 mm}
\noindent \textbf{AUTHOR CONTRIBUTIONS} \\
\noindent K.R.A. wrote the manuscript and created the figures.

\vspace{4 mm}
\noindent \textbf{COMPETING INTERESTS} \\
\noindent The author declares no competing financial or non-financial interests.

\bibliography{control_library}
\end{document}